\documentclass[a4paper,11pt,twoside]{article}
\pdfoutput=1

\usepackage{geometry} 
\usepackage{a4wide}

\usepackage{amssymb,amsmath,bm}
\usepackage{cite}
\usepackage{url}
\usepackage{color}
\usepackage{slashed}
\usepackage{graphicx}
\usepackage[latin1]{inputenc}
\usepackage{amsfonts}
\usepackage{lscape}
\usepackage{amsthm}
\usepackage{booktabs}
\usepackage{array}
\usepackage{rotating}
\usepackage{float}
\usepackage{xspace}
\usepackage{accents}

\usepackage{mathrsfs}

\usepackage[small,bf]{caption}
\selectfont

\begin{document}

\begin{titlepage}


\begin{center}
{
\bf\huge
Low Fine Tuning in Yukawa-deflected \\ \vspace{8pt}
Gauge Mediation
}
\\[10mm]
{\Large Waqas~Ahmed$^{\,\star}$} \footnote{E-mail: \texttt{waqasmit@itp.ac.cn}},
{\Large Lorenzo~Calibbi$^{\,\star}$} \footnote{E-mail: \texttt{calibbi@itp.ac.cn}},
{\Large Tianjun~Li$^{\,\star\,\heartsuit \,\dagger}$} \footnote{E-mail: \texttt{tli@itp.ac.cn}},\\[1mm]
{\Large Azar~Mustafayev$^{\,\ddag}$} \footnote{E-mail: \texttt{mustafayev@physics.umn.edu}},
{\Large Shabbar~Raza$^{\,\ast}$} \footnote{E-mail: \texttt{shabbar.raza@fuuast.edu.pk}}\\[1mm]
\end{center}
\vspace*{0.50cm}
\centerline{$^{\star}$ \it
CAS Key Laboratory of Theoretical Physics and Kavli Institute for Theoretical Physics}
\centerline{\it
China (KITPC), Institute of Theoretical Physics, Chinese Academy of Sciences,}
\centerline{\it Beijing 100190, P.~R.~China}
\vspace*{0.2cm}
\centerline{ $^\heartsuit $\it 
School of Physical Sciences, University of Chinese Academy
of Sciences,} 
\centerline{\it  Beijing 100049, P. R. China}
\vspace*{0.2cm}
\centerline{$^{\dagger}$ \it
School of Physical Electronics, University of Electronic Science and Technology of China,}
\centerline{\it Chengdu 610054, P.~R.~China}
\vspace*{0.2cm}
\centerline{$^{\ddag}$ \it
William I.~Fine Theoretical Physics Institute, University of Minnesota,}
\centerline{\it
Minneapolis, MN 55455, USA}
\vspace*{0.2cm}
\centerline{$^{\ast}$ \it
Department of Physics, Federal Urdu University of Arts, Science and Technology,}
\centerline{\it Karachi 75300, Pakistan}

\vspace*{1.cm}

\begin{abstract}
\noindent 
We discuss a class of models with gauge-mediated supersymmetry breaking
characterized by a non-unified messenger sector inducing non-standard gaugino mass ratios, 
as well as by additional contributions to the soft mass terms from a matter-messenger coupling.
The well-known effect of this coupling is to generate A-terms at one-loop level, hence raising
the Higgs mass without relying on super-heavy stops. At the same time, a hierarchy between Wino
and gluino masses, as induced by the non-unified messenger fields, can greatly lower the radiative corrections
to the Higgs soft mass term driven by the high-energy parameters, thus reducing the fine tuning.
We search for models with low fine tuning within this scenario, and we discuss the spectrum, 
collider phenomenology, constraints, and prospects of the found solutions. 
We find that some setups are accessible or already excluded by searches at the Large Hadron Collider, 
and all our scenarios with a tuning better than about 2\% can be tested at the International Linear Collider.
\end{abstract}

\end{titlepage}

\addtocounter{footnote}{-5}

\section{Introduction}
\label{sec:intro}
Despite the numerous dedicated searches, the first years of run of the Large Hadron Collider (LHC)
have given no sign of new phenomena associated 
with supersymmetry (SUSY). The stringent limits set by ATLAS and CMS on the SUSY spectrum, as well as the measured Higgs boson mass, $m_h \simeq 125$ GeV, certainly challenge the paradigm of low-energy SUSY as a solution of the electro-weak hierarchy problem, especially within the simplest realizations, such as minimal supergravity, minimal gauge mediation (mGMSB) etc.
For this reason, we find it timely to study less minimal constructions in search for realizations that can better accommodate (i) the observed Higgs mass, and (ii) a spectrum heavy enough to evade the direct LHC searches, but (iii) without a large fine-tuning (FT) among the theory's parameters. 

Although any attempt to quantify the FT price of a model -- which is obviously not a physical observable -- is to some extent subjective, we still regard the FT measures proposed in the literature, starting from \cite{Barbieri:1987fn}, as useful tools to compare different models and SUSY spectra, and identify those that are most likely realized in nature.
 
In a previous work \cite{Calibbi:2016qwt}, we employed the ``electro-weak FT'', $\Delta_{\rm EW}$, introduced in \cite{Baer:2012up,Baer:2012mv}, in order to show that models with gauge-mediated SUSY breaking \cite{Giudice:1998bp} with messengers in incomplete representations of a grand unified gauge group can lower the FT by more than one order of magnitude with respect to mGMSB, i.e.~$\Delta_{\rm EW}\lesssim 50$. 
In particular, solutions with a reduced FT are possible if such non-unified messenger sector gives raise to a Wino mass substantially larger than the gluino mass at the mediation scale. This in turn induces a compensation between gauge and Yukawa radiative corrections to the Higgs soft mass $\widetilde{m}^2_{H_u}$, reducing its sensitivity to stop and gluino masses. 
This effect can be appreciated by inspecting the renormalization group equation (RGE) of $\widetilde{m}^2_{H_u}$, 
whose one-loop expression is:
\begin{equation}
16\pi^2 \frac{d}{dt} \widetilde{m}^2_{H_u} \approx 6 y_t^2 \left[\widetilde{m}^2_{H_u} + \widetilde{m}^2_{Q_3}+ \widetilde{m}^2_{U_3}\right] + 6 A_t^2 - 6g_2^2M_2^2\quad \left( t\equiv \log\frac{\mu}{M} \right)\,,
\label{eq:rge}
\end{equation}
where $\mu$ is the renormalization scale and $M$ a reference scale, and we omitted the hypercharge-dependent terms. 
The terms proportional to the top A-term $A_t$ and the top Yukawa $y_t$ (where the left-handed (LH) and right-handed (RH) stop masses
 $\widetilde{m}^2_{Q_3}$ and $\widetilde{m}^2_{U_3}$ appear) have opposite sign with respect to the $SU(2)$ gauge term 
 (where $g_2$ is the gauge coupling and $M_2$ the gaugino mass). As a consequence a compensation between the two terms,
 hence a reduced value of $|\widetilde{m}^2_{H_u}|$ at low energies, is possible provided that $M_2>M_3$ (as the gluino mass $M_3$
 induces large positive contributions to the stop masses in the running). For related works, see \cite{Brummer:2012zc,Brummer:2013yya,Bhattacharyya:2015vha,Fukuda:2015pra,Gogoladze:2016grr,Gogoladze:2016jvm}.
 
In \cite{Calibbi:2016qwt}, we found that the typical spectra of these low-tuned models with $M_2>M_3$ tend to lie in the multi-TeV range (mainly because of $m_h$), and the only sub-TeV states are the Higgsino and possibly Bino and right-handed sleptons. The absence of signals at the LHC searches is therefore a natural consequence of the framework. However, this also makes it very difficult to test even in the long run. A similar conclusion is shared in the context of gravity mediation by grand unified theory (GUT) models
with the non-universal gaugino masses induced by the breaking of the GUT group \cite{Gogoladze:2009bd,Horton:2009ed,Antusch:2012gv,Gogoladze:2012yf,Gogoladze:2013wva,Kaminska:2013mya,Martin:2013aha,Kowalska:2014hza}.

The aim of the present work is to verify whether a deformation of models with non-unified messenger sectors can modify the above conclusion, and investigate possible handles to test this class of low-tuned models at the LHC. 
For this purpose, we introduce in the Lagrangian a Yukawa-like coupling between messenger and matter superfields, which provides additional contributions to the soft terms besides those purely due to gauge interactions.  
Models of this kind -- sometimes labelled as ``Yukawa-deflected gauge mediation'' or ``Extended gauge mediation'' -- have been proposed long ago \cite{Dine, Chacko1, Chacko2}, and more recently received renewed attention \cite{Shadmi:2011hs,Y1, Jelinski:2011xe, Y2, Kang, Craig, Babu, Shadmi2, Ray, Craig:2013wga,Evans:2013kxa,CPZ1,Jelinski:2013kta,Galon:2013jba,Ding:2013pya,CPZ2,Allanach:2015cia,Evans:2015swa,Jelinski:2015voa,Ierushalmi:2016axs,Allanach:2016pam,Byakti:2016ayi,Kang:2016iok,Everett:2016meb}, especially after the discovery of the Higgs, since they can induce one-loop contributions to the stop A-term $A_t$ -- in contrast to ordinary gauge mediation setups -- such that $m_h \simeq 125$ GeV can be accommodated without paying the price of multi-TeV stop masses. Despite that, it has been shown that such models typically do not improve much the FT over mGMSB \cite{Craig:2013cxa,Casas:2016xnl}, because the messenger-matter coupling does not only generate $A_t$, but also large negative contributions to $\widetilde{m}^2_{H_u}$ of the order $\approx-|A_t|^2$, so that large cancellations among the parameters in the Higgs potential
are still needed in order to achieve a correct electro-weak symmetry breaking (EWSB).
However, to the best of our knowledge, 
marrying ``Yukawa deflection'' with a non-unified messenger sector has not been attempted yet.  
In this extended setup, we expect that -- similarly to the case of our previous work -- negative contributions to the Higgs soft masses can be compensated by a heavy Wino in the renormalization group running, reducing the amount of cancellation required by EWSB, i.e.~the fine tuning. At the same time, the $A_t$ induced by the matter-messenger coupling should allow to obtain $m_h \simeq 125$ GeV with a $\mathcal{O}(1)$ TeV stop, unlike in \cite{Calibbi:2016qwt}. 

The rest of the paper is organized as follows. In section \ref{sec:model}, we present the model setup and the high-energy parameters that control the soft SUSY-breaking mass terms. In section \ref{sec:results}, we show the results of a numerical scan
over the space of models, highlighting the solutions with reduced fine tuning and the typical features of their SUSY spectra. 
We discuss the LHC phenomenology in section \ref{sec:lhc} of the different classes of low-tuned solutions we found, in particular present collider constraints, and testability prospects. We draw our conclusions in section \ref{sec:conclusions}.

\section{Model setup and soft masses}
\label{sec:model}
As in \cite{Calibbi:2016qwt}, we work in the context of the minimal supersymmetric standard model (MSSM) and 
we consider a Gauge Mediation setup where SUSY breaking is transmitted 
to the visible sector trough loops involving heavy messengers in vectorlike representations of the Standard Model (SM) gauge group, 
which however do not necessarily belong to complete multiplets of a grand unified group. 
The resulting contributions to gaugino and sfermion masses then depend on three independent parameters $b_1^M,~b_2^M,~b_3^M$, denoting the shifts induced at the messenger scale $M$ to the 1-loop $\beta$-function coefficients of the SM gauge couplings
$g_1,~g_2,~g_3$.\footnote{A discussion of the possible sets of messengers and the resulting $b_a^M$ has been presented in \cite{Calibbi:2016qwt}.}
As in minimal Gauge Mediation, gaugino masses ($M_a$, $a=1,2,3$) arise at 1-loop, scalar masses ($\widetilde{m}^2_X (M)$, $X=Q,\,U,\,D,\,L,\,E,\,H_u\,H_d$) at 2-loops. Their expressions for non-unified messengers read at the messenger scale \cite{Martin:1996zb}:
\begin{align}
\label{eq:gauginos}
M_a (M) &= \frac{\alpha_a(M)}{4\pi} b_a^M \Lambda,~~a=1,2,3, \\
\widetilde{m}^2_X (M) &= 2\sum_{a=1,3} \left(\frac{\alpha_a(M)}{4\pi}\right)^2 C^X_a b_a^M \Lambda^2,
\label{eq:scalars}
\end{align}
where $\Lambda \equiv F/M$ is the ratio of a single SUSY-breaking F-term and the mediation scale $M$, 
and $C^X_a$ ($a=1,2,3$) is the quadratic Casimir of the representation of $X$ under $SU(3)\times SU(2) \times U(1)$.
Being of a purely gauge origin, the sfermion masses are flavor universal. 

In the spirit of `Yukawa-deflected' gauge mediation, we also allow for matter-messenger couplings.
Among several possibilities \cite{Evans:2013kxa}, we choose to introduce a single new coupling involving only the third generation quarks, of the form:
\begin{equation}
W \supset \lambda_t \,Q_3 U_3 \Phi_u,
\end{equation}
where $\Phi_u$ is a messenger superfield with the same quantum numbers as $H_u$.

Unlike in mGMSB, squark A-terms are generated at 1-loop and read the messenger scale \cite{Y1,Evans:2013kxa}:
\begin{equation}
\label{eq:yukdef1}
A_t(M) = -\frac{3\Lambda}{16\pi^2} \lambda_t^2 y_t ,\quad A_b (M)= -\frac{\Lambda}{16\pi^2} \lambda_b^2 y_b,
\end{equation}
where $y_t$ and $y_b$ are the ordinary top and bottom Yukawas.

Additionally, negative 1-loop contribution to the stop masses are generated \cite{Y1,Evans:2013kxa}: 
\begin{equation}
\Delta {\widetilde{m}}^{2\,(1)}_{Q_3}= -\frac{\Lambda^2}{96\pi^2} \lambda_t^2 ~g(x),\quad 
\Delta {\widetilde{m}}^{2\,(1)}_{U_3}= -\frac{\Lambda^2}{48\pi^2} \lambda_t^2 ~g(x),
\end{equation}
with
\begin{equation}
g(x)= 3 \frac{(x-2)\log(1-x)-  (2+x) \log(1+x)}{x^2}  = x^2 + \frac{4}{5}x^4 + \mathcal{O}({x^6}),\quad x \equiv \frac{\Lambda}{M}.
\end{equation}
These contributions become irrelevant for $M\gg \Lambda$.

Finally, third generation squarks and the two Higgs soft masses are deflected by additional 2-loop contributions that do not vanish 
for large messenger scales \cite{Y1,Evans:2013kxa}:
\begin{align}
\label{eq:stopL}
\Delta {\widetilde{m}}^{2\,(2)}_{Q_3} &= \frac{\Lambda^2}{256\pi^4} \left[ -\left(\frac{13}{15}g_1^2 + 3g_2^2 +\frac{16}{3} g_3^2\right)\lambda_t^2 +6 \lambda_t^4 + 6 \lambda_t^2 y_t^2  \right],\\
\label{eq:stopR}
\Delta {\widetilde{m}}^{2\,(2)}_{U_3} &= \frac{\Lambda^2}{128\pi^4} \left[ -\left(\frac{13}{15}g_1^2 + 3g_2^2 +\frac{16}{3} g_3^2\right)\lambda_t^2 +6 \lambda_t^4 + 6 \lambda_t^2 y_t^2  + \lambda_t^2 y_b^2 \right],\\
\Delta {\widetilde{m}}^{2\,(2)}_{D_3} &= -\frac{\Lambda^2}{128\pi^4}  \lambda_t^2 y_b^2 ,\\
\Delta {\widetilde{m}}^{2\,(2)}_{H_u} &= -\frac{9 \Lambda^2}{256\pi^4}  \lambda_t^2 y_t^2 ,\label{eq:m2hu}\\
\Delta {\widetilde{m}}^{2\,(2)}_{H_d} &= -\frac{3\Lambda^2}{256\pi^4}  \lambda_t^2 y_b^2 .
\label{eq:yukdef2}
\end{align}
Unlike the standard contributions to the sfermion masses in Eq.~(\ref{eq:scalars}), these additional terms induced by
$\lambda_t$ only concern the stops, thus they provide a departure from flavor universality, hence from the Minimal
Flavor Violation (MFV) \cite{D'Ambrosio:2002ex} structure of low-energy squark mass matrices that is characteristic of pure GM frameworks.
However, it has been shown that this does not induce unacceptably large flavor-changing neutral current processes,
as long as the matter-messenger couplings feature a flavor hierarchical structure resembling that of the ordinary
Yukawa interactions \cite{Shadmi:2011hs,CPZ1,CPZ2}. 
This is trivially the case in the present setup, as we introduced only a third-generation
coupling  $\lambda_t$ of $\mathcal{O}(1)$.

\section{Models with low tuning and typical spectra}
\label{sec:results}
In order to explore the class of models defined in the previous section, we performed a numerical scan employing 
a version of the routine {\tt ISAJET 7.85} \cite{Baer:1999sp} that we modified implementing the Yukawa-deflection
contributions to the soft terms given in Eqs.~(\ref{eq:yukdef1})-(\ref{eq:yukdef2}).
As in \cite{Calibbi:2016qwt}, the messenger contributions to the $\beta$-function coefficients $b_1^M,~b_2^M,~b_3^M$ are integer numbers that
we randomly varied  within these intervals:
\begin{align}
1 \le~ (5\times b_1^M) \le 75, \quad 1 \le ~ b_2^M \le 20, \quad 1 \le ~ b_3^M \le 7.
\label{eq:branges}
\end{align}
For the other parameters we took the following ranges:
\begin{align}
 5\times10^4~{\rm GeV} \le  ~\Lambda \le 10^6 ~{\rm GeV},&\quad 2\times \Lambda \le  ~M   \le 10^{15} ~{\rm GeV}, \nonumber\\
  5 \le  ~\tan\beta ~ \le 50,&\quad 0 \le ~\lambda_t \le 1.5. & 
\label{eq:ranges}
\end{align}
\begin{figure}[t!]
\begin{center}
\includegraphics[width=0.47\textwidth]{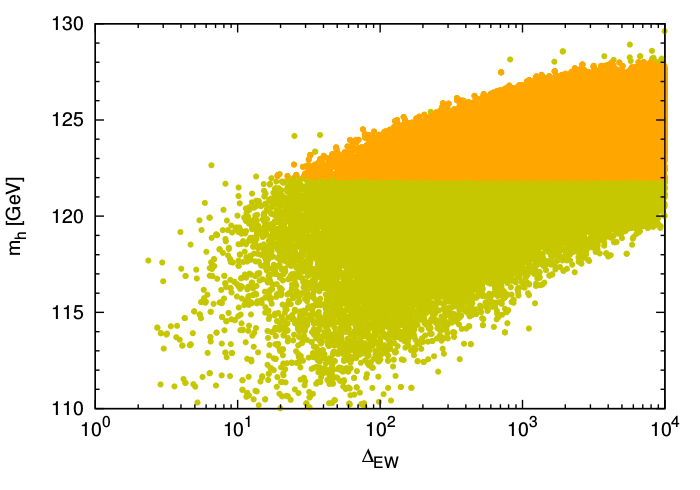}
\hfill
\includegraphics[width=0.47\textwidth]{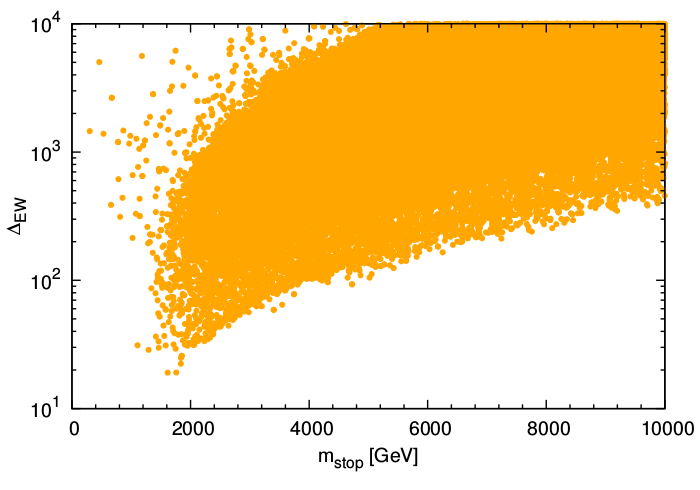}\\
\includegraphics[width=0.47\textwidth]{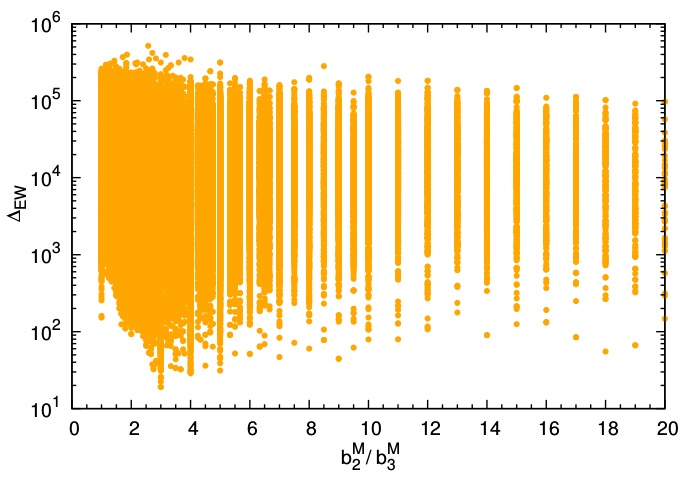}
\hfill
\includegraphics[width=0.47\textwidth]{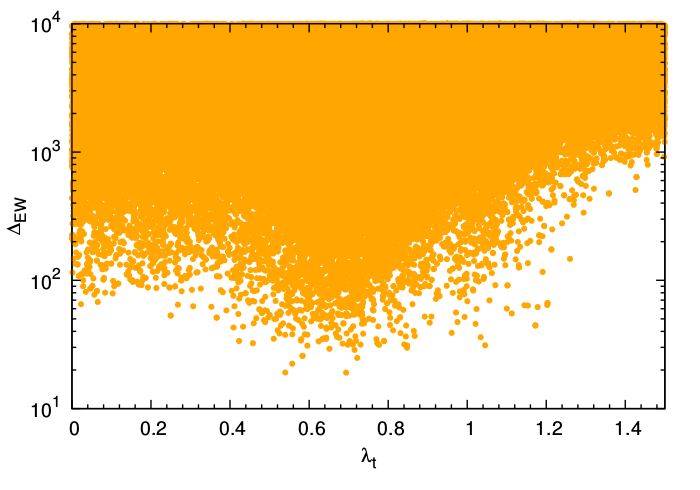}
\end{center}
\caption{First line: $\Delta_{\rm EW}$ vs.~$m_h$ (left), the lightest stop mass (right). Second line
$\Delta_{\rm EW}$ vs.~the ratio $b_2^M/b^M_3$ (left), $\lambda_t$ (right). Orange points correspond to
$122~{\rm GeV} \le  m_h \le 128~{\rm GeV}$.
\label{fig:mh-FT} }
\end{figure}

In the following, we base our naturalness considerations on the `electro-weak' fine tuning, defined as \cite{Baer:2012up,Baer:2012mv}
\begin{equation}
\Delta_{\rm EW} \equiv \frac{\max_x |C_x|}{m^2_Z/2},
\label{eq:DeltaEW}
\end{equation}
where $C_x$ are the terms in the right-hand side of the minimization condition of the Higgs potential:
\begin{equation}
\frac{m_Z^2}{2} = \frac{(\widetilde{m}^2_{H_d} +\Sigma_d) - (\widetilde{m}^2_{H_u} +\Sigma_u)\tan^2\beta}{\tan^2\beta-1} - \mu^2.
\label{eq:ewsb}
\end{equation}
The quantities $\Sigma_{u,d}$ express the 1-loop corrections to the tree-level potential~\cite{Baer:2012cf}. 

We show in Fig.~\ref{fig:mh-FT} the resulting $\Delta_{\rm EW}$ for the points of our scan, as defined by Eqs.~(\ref{eq:branges}, \ref{eq:ranges}). The orange points correspond to the observed Higgs mass, once a theoretical uncertainty of 3 GeV is
taken into account: $122~{\rm GeV} \le  m_h \le 128~{\rm GeV}$. As we can see, this condition is satisfied by points
with $\Delta_{\rm EW}$ as low as $\approx 20$, which corresponds to a tuning of about 5\%,
cf.~the top-left plot of Fig.~\ref{fig:mh-FT}.
\begin{figure}[t]
\begin{center}
\includegraphics[width=0.47\textwidth]{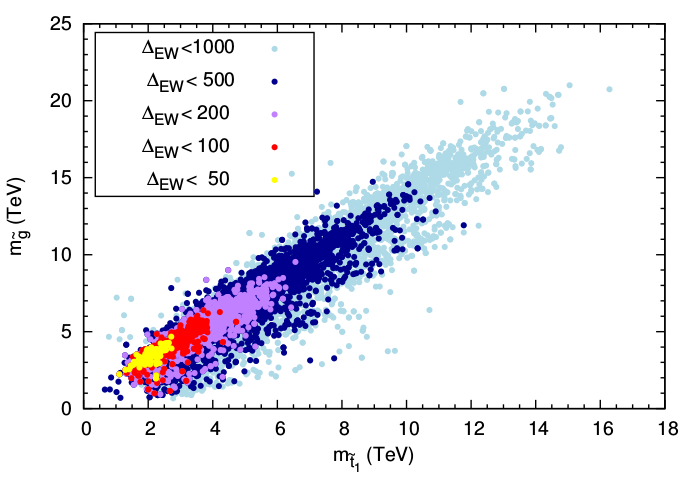}
\hfill
\includegraphics[width=0.47\textwidth]{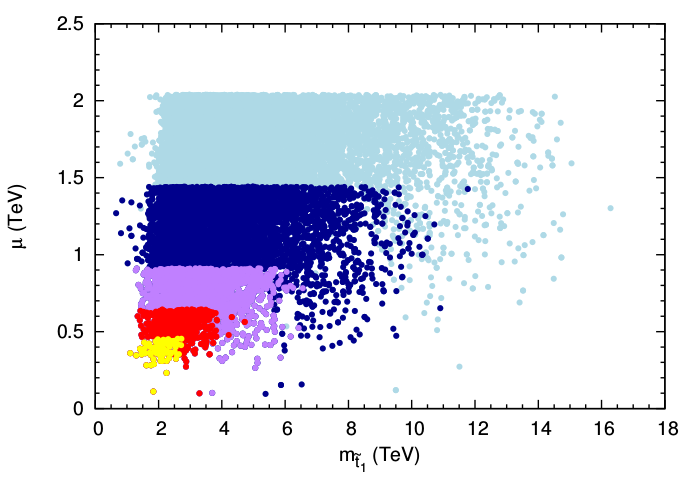}
\caption{Gluino (left panel) and Higgsino (right panel) 
mass vs.~lightest stop mass for different ranges of $\Delta_{\rm EW}$. \label{fig:spectrum1} }
\end{center}
\end{figure}
From the expression for $\Delta_{\rm EW}$, Eq.~(\ref{eq:DeltaEW}), it is clear that models with reduced tuning will require in particular that $|\widetilde{m}^2_{H_u}|$ and $|\Sigma_u|$ are not much larger than $m_Z^2$. The first condition is facilitated if the messenger sector is such that $b^M_2 >b^M_3$, i.e.~the Wino mass exceeds the gluino mass at the messenger scale, as shown in \cite{Calibbi:2016qwt}. In fact, this can lead to a compensation
of the terms $\propto y_t^2 \left(\widetilde{m}^2_{Q_3}+ \widetilde{m}^2_{U_3} \right)$ and $\propto g_2^2M_2^2$ in the
$\beta$-function of $\widetilde{m}^2_{H_u}$, cf.~Eq.~(\ref{eq:rge}), 
reducing the sensitivity of the low-energy value of $|\widetilde{m}^2_{H_u}|$ on 
heavy gluinos and stops. This mechanism is particularly efficient for $b^M_2 \approx 3\times b^M_3$ as shown
in the bottom-left panel of Fig.~\ref{fig:mh-FT}.\footnote{In models with $\lambda_t=0$, $b^M_2/b^M_3$ can not be larger than
about 2.5, a value that provides the solutions with best FT \cite{Calibbi:2016qwt}, because of failure of EWSB for a too large
positive contribution of $M_2$ to the running of $\widetilde{m}^2_{H_u}$. This effect is compensated for
$\lambda_t \neq0$ by the additional negative contribution in Eq.~(\ref{eq:m2hu}), and large ratios $b^M_2/b^M_3$
are accessible.}
A very heavy stop sector would however reintroduce a fine tuning problem inducing large finite radiative corrections
encoded in $\Sigma_u$. 
The additional contributions induced by the coupling $\lambda_t$ allow for solutions with a light stop,
which is a qualitatively different with respect to the models with $\lambda_t =0$ considered in our previous work. 
This is shown in the top-right panel of Fig.~\ref{fig:mh-FT}, where we we see that points with $\Delta_{\rm EW}\lesssim 50$
can feature $m_{\widetilde{t}_{1}}$ down to 1 TeV.
This is due to different effects: the new contributions to the stop soft masses in Eqs.~(\ref{eq:stopL}, \ref{eq:stopR})
are negative for $\lambda_t \lesssim 0.7$ \cite{CPZ1}; a sizable $A_t$ (generated at 1-loop $\propto \lambda_t^2$) 
allows for $m_h \approx 125$ GeV with a lighter stop sector and, at the same time, lowers the mass of the lightest stop eigenstate through a large LR stop mixing. As a result, we find the lowest values of $\Delta_{\rm EW}$ for $\lambda_t \approx 0.6\div 0.8$,
as shown in the bottom-right plot of Fig.~\ref{fig:mh-FT}.
\begin{figure}[t]
\begin{center}
\includegraphics[width=0.47\textwidth]{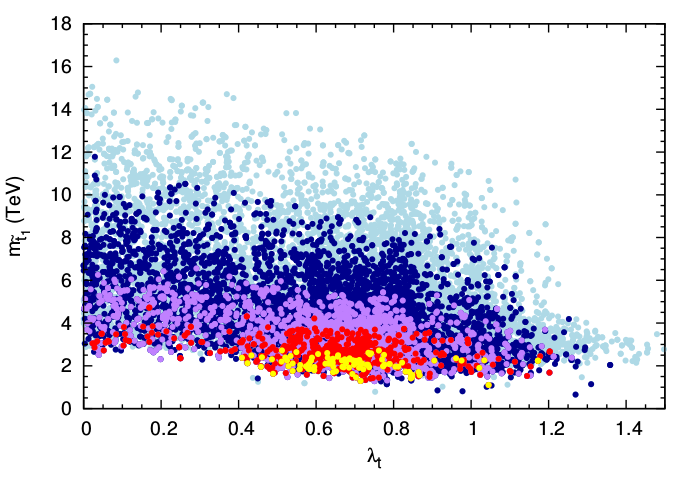}
\hfill
\includegraphics[width=0.47\textwidth]{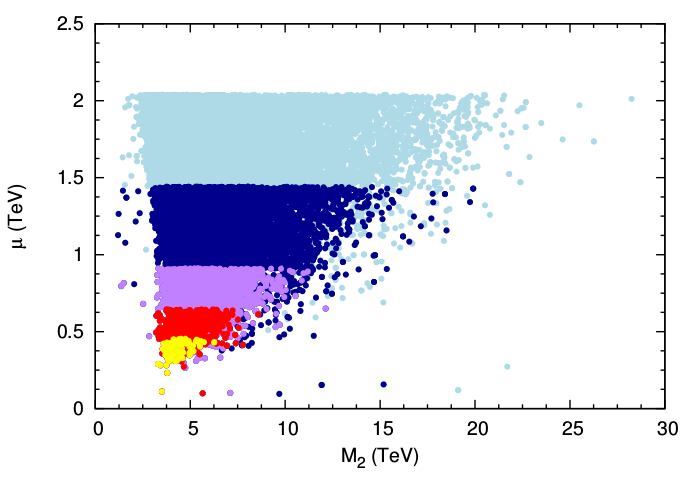}
\caption{Stop mass vs.~the matter-messenger coupling $\lambda_t$ (left panel); Higgsino mass vs.~Wino mass (right panel). 
Color code as in Fig.~\ref{fig:spectrum1}.
\label{fig:spectrum2} }
\end{center}
\end{figure}
\begin{figure}[t]
\begin{center}
\includegraphics[width=0.47\textwidth]{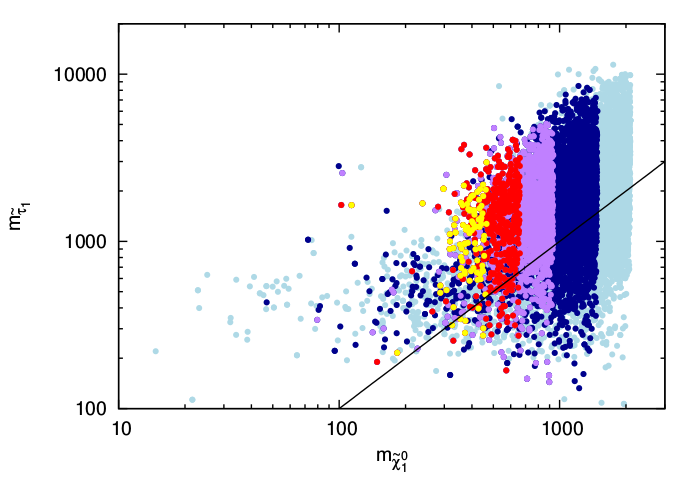}
\hfill
\includegraphics[width=0.47\textwidth]{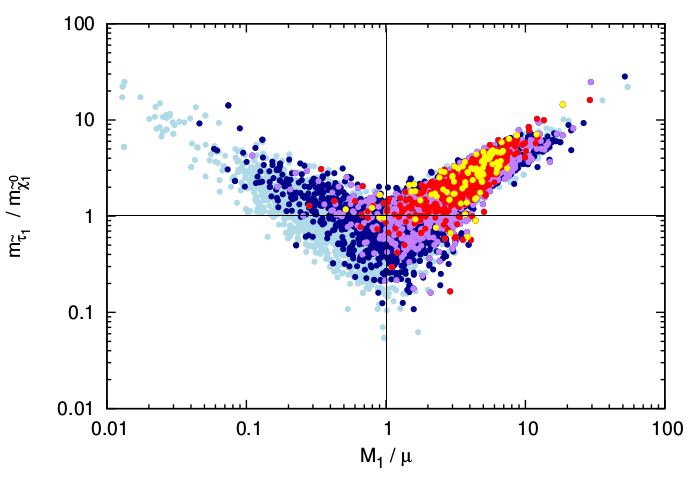}
\caption{Lightest stau vs.~lightest neutralino mass (left panel); stau-neutralino mass ratio vs.~Bino-Higgsino mass ratio (right panel).  Color code as in Fig.~\ref{fig:spectrum1}.\label{fig:spectrum3} }
\end{center}
\end{figure}

Characteristic features of the spectrum of solutions with low fine tuning and  $122~{\rm GeV} \le  m_h \le 128~{\rm GeV}$
can be seen in Figs.~\ref{fig:spectrum1}-\ref{fig:spectrum3}, where different ranges of $\Delta_{\rm EW}$ are plotted in different colors. From these plots we see that the spectrum of the solutions with lowest $\Delta_{\rm EW}$ is characterized  
by the following mass ranges:
\begin{align}
 1~{\rm TeV}\lesssim ~ &m_{\widetilde{t}_1} ~\lesssim 2.5~(4.5)~{\rm TeV}\,, \nonumber \\
 \Delta_{\rm EW} \lesssim 50~(100) \quad \Rightarrow  \quad  
 2~(1)~{\rm TeV}~ \lesssim ~ &m_{\widetilde{g}}~ \lesssim 4.5~(6.5)~{\rm TeV}\,, \\
100~{\rm GeV }\lesssim ~ &m_{\widetilde{\chi}^0_{1}}  ~\lesssim 450~(650)~{\rm GeV }\,. \nonumber
\end{align}
Furthermore, we see that  solutions $\Delta_{\rm EW} < 50$ (yellow points) have quite heavy Winos at low energy ($M_2\gtrsim 3.5$ TeV), 
which reflects the high-energy condition on gaugino masses, as explained above, and a sizable Yukawa deflection 
is preferred, $\lambda_t \gtrsim 0.4$, cf.~Fig.~\ref{fig:spectrum2}.

We are particularly interested in the nature of the next to lightest SUSY particle (NLSP) (the lightest SUSY particle (LSP) is always a light gravitino, as typical of Gauge Mediation), as it determines most of the collider phenomenology 
of our models, as we will discuss in the next section. 
Information about the NLSP can be read in Fig.~\ref{fig:spectrum3}.
Our typical low-tuned solutions, e.g. $\Delta_{\rm EW}< 50$ (yellow points) and $\Delta_{\rm EW}< 100$ (red points),
feature as NLSP a neutralino that is mostly Higgsino: indeed, the corresponding points typically 
lie in the top-right quadrant of the second plot. However, we also found some solutions with a Bino-like NLSP (top-left
quadrant) and stau NSLP (bottom quadrants). 

\begin{table}[t!]
{\footnotesize
\begin{center}
\begin{tabular}{@{} |c||c|c|c||c|c|c| @{}}
\hline    \hline    
     & {\bf A1} & {\bf A2} & {\bf A3} & {\bf B1} & {\bf B2} & {\bf B3} \\ 
\hline    
    $\Lambda$ & 91 TeV & 107 TeV & 80 TeV & 131 TeV   & 144 TeV  & 184 TeV \\ 
    $M$   & $2 \times10^{14}$ & $8 \times10^{14}$ & $2 \times10^{14}$  & $5\times10^{13}$ & $2.5\times10^{11}$  &  $2.0\times10^{6}$   \\ 
    $\tan\beta$ & 31 & 17  & 20 &  29  &18 & 24  \\ 
    $(b^M_1,~b^M_2,~b^M_3)$ & (7, 15, 5) & ($\frac{18}{5}$, 15, 3) & ($\frac{73}{5}$, 19, 7) & ($\frac{23}{5}$, 14, 5)  & (1, 14, 4) & 
    ($\frac{38}{5}$, 16, 3)    \\ 
    $\lambda_t$ & 0.69 & 1.05  & 0.54 & 0.59  & 0.76 & 0.89  \\ 
\hline    
    ${\Delta_{\rm EW}}$ & {\bf 19} & {\bf 31}  & {\bf 38} & {\bf 38} &{\bf 44} & {\bf 90}    \\ 
    ${\Delta_{\rm BG}}$  {w/o} ${\lambda_t}$ & {\bf 62} & {\bf 51} & {\bf 50} & {\bf 109}  & {\bf 53} & {\bf 80}   \\ 
         $\Delta_{\rm BG}$ wrt $\lambda_t$ & 214 & 3064 & 138 & 567  & 1845 & 4174 \\ 
         $\lambda_t\times\Lambda$ & 49 TeV & 112 TeV & 43 TeV & 77 TeV & 109 TeV & 164 TeV \\
    \hline    
        $\mu$   & 289 & 360 & 395 & 364 & 360 & 578   \\ 
        $m_h,~m_A$   & 122, 2366 & 122, 3206 & 122, 2863 & 123, 3147  & 122, 3163 & 123, 3255 \\ 
        \hline    
        $m_{\widetilde{g}}$   & 3018 & 2231 & 3638 & 4227  & 3776 & 3705   \\ 
    $m_{\widetilde{\chi}^0_{1,2}}$   & \textcolor{red}{287}, 291 &  \textcolor{red}{362}, 370 &  \textcolor{red}{405}, 408 & 372, 376  
  &  \textcolor{red}{183}, 383      &  \textcolor{red}{594}, 596 \\ 
     $m_{\widetilde{\chi}^0_{3,4}}$   & 870, 3443 & 531, 4071 & 1602, 3817 & 825, 4625  & 385, 5061 & 1950, 7389  \\ 
    $m_{\widetilde{\chi}^\pm_{1,2}}$   & 299, 3431 & 381, 4069 &  419, 3800& 387, 4606  & 394, 5047 & 613, 7374   \\     
    $m_{\widetilde{t}_{1,2}}$  & {\bf 1615}, 3395 & {\bf 1107}, 3629 & {\bf 2109}, 3909 & {\bf 2381}, 4638  &  {\bf 2512}, 4403 &  {\bf 3552}, 5611\\     
    $m_{\widetilde{b}_{1,2}}$  & 2473, 3419 & 1974, 3672 & 3119, 3928 & 3539, 4667  & 3329, 4419 & 3385, 4815 \\     
    $m_{\widetilde{\tau}_{1,2}}$  & 493, 2707 & 763, 3292 & 1280, 3003 &\textcolor{red}{283}, 3537 & 216, 3257  & 990, 3368  \\     
    $m_{\widetilde{u}_L},\,m_{\widetilde{d}_L}$  & 3807, 3808 & 3849, 3850 & 4314, 4314 & 5155, 5156  & 4679, 4674 & 4777, 4778    \\     
     $m_{\widetilde{u}_R},\,m_{\widetilde{d}_R}$  & 2767, 2722 & 2038, 2077 & 3324, 3226 & 3830, 3795 & 3385, 3397 & 3486, 3458 \\     
    $m_{\widetilde{\ell}_L},\,m_{\widetilde{\ell}_R}$  & 2762, 916 & 3304, 903 & 3028, 1390 & 3595, 919 & 3271, 362 & 3391, 1069  \\     
  \hline
  NLSP &  $\widetilde{\chi}^0_{1}$ ($\approx \widetilde H$) & $\widetilde{\chi}^0_{1}$ ($\approx \widetilde H$) &
   $\widetilde{\chi}^0_{1}$ ($\approx \widetilde H$) &  $\widetilde{\tau}_{1}$ ($\approx \widetilde{\tau}_{R} $)  & $\widetilde{\chi}^0_{1}$ ($\approx \widetilde B$) &    $\widetilde{\chi}^0_{1}$ ($\approx \widetilde H$)\\
   $m_{\widetilde G}$  & 4.3 GeV & 20 GeV & 3.8 GeV & 1.6 GeV  & 8.6 MeV  & 87 eV  \\ 
  $c \tau_{\rm NLSP}$ (m) &$1.2\times10^{13}$ & $6.4\times10^{13}$ & $1.2\times10^{12}$  & $4.7\times10^{11}$    &   $1.5\times10^{8}$ & $8.3\times10^{-5}$\\
 \hline  \hline    
  \end{tabular}
 \end{center}
 }
  \caption{Spectrum and parameters of six representative models. Models  {\bf A1},  {\bf A2}, and  {\bf A3} belong to the 
 most  typical class of low-tuned solutions, those featuring an almost pure (long-lived) Higgsino NLSP.
 {\bf A1} and  {\bf A2} correspond respectively to the model with lowest tuning and lightest $\widetilde{t}_1$ we found in the scan. 
Model  {\bf A3} is characterized by a reduced sensitivity on $\lambda_t$, due to a small value of $\lambda_t\times\Lambda$, see the text for details. Models  {\bf B1},  {\bf B2}, and  {\bf B3} illustrate particular corners of the parameter space. 
Model  {\bf B1} is an example of the (long-lived) stau NLSP scenario. 
 Model  {\bf B2} features a Bino-like neutralino NLSP.
  Model  {\bf B3} show a peculiar solution with a short-lived (promptly-decaying) Higgsino NLSP.
    Dimensionful quantities are in GeV unless otherwise indicated. 
\label{tab:benchmarks}}
\end{table}

The above possibilities are exemplified by the benchmark models whose spectra are listed in Tab.~\ref{tab:benchmarks}.
Models  {\bf A1}-{\bf A3} are examples of the typical low-tuned setup with a rather light Higgsino NLSP. 
In particular, model  {\bf A1} corresponds to the solution with the lowest $\Delta_{\rm EW}$ we found in the scan, and illustrates a typical spectrum with Higgsino NLSP and heavy spectrum, with the $SU(3)$ and $SU(2)$ singlets as the only other states possibly lighter than 1 TeV. Model  {\bf A2} also features a Higgsino NSLP, but lighter stop and gluino masses (around 1.1 and 2.2 TeV respectively). 

Besides  ${\Delta_{\rm EW}}$, we also computed the Barbieri-Giudice FT measure  ${\Delta_{\rm BG}}$ \cite{Barbieri:1987fn}
for the benchmark models:
\begin{equation}
\Delta_{\rm BG} \equiv \max \Delta_{\rm BG}(A),\quad  \Delta_{\rm BG}(A)= \left|\frac{\partial \log m_Z^2}{\partial \log A}\right|,
\end{equation}
where $A$ run over the fundamental high-energy parameters: $\Lambda$, $M$, $\mu^2$, and the matter-messenger
coupling $\lambda_t$. In Tab.~\ref{tab:benchmarks}, we display $\Delta_{\rm BG}$ both with and without taking into account
$\Delta_{\rm BG}(\lambda_t)$. As we can see, the resulting $\Delta_{\rm BG}$ is of the same order as $\Delta_{\rm EW}$, 
if we do not consider $\Delta_{\rm BG}(\lambda_t)$, while this latter quantity can be considerably larger. This is not surprising
given the large negative Yukawa-deflection contribution to ${\widetilde m}^2_{H_u}$, cf.~Eq.~(\ref{eq:m2hu}). Such a sensitivity
on $\lambda_t$ is in fact larger for scenarios with higher values of $\lambda_t \times \Lambda$ (that we also show in the Table),
since this is the quantity that actually controls the size of the Yukawa-deflected contributions to the soft masses,
cf.~Eqs.~(\ref{eq:stopL}-\ref{eq:yukdef2}), in particular to ${\widetilde m}^2_{H_u}$.
We can argue that indeed $\Delta_{\rm BG}(\lambda_t)$ should not be considered in computing the FT (similarly to
$\Delta_{\rm BG}(y_t)$), since $\lambda_t$ is a Yukawa coupling. 
In fact, assuming that the MSSM arise from string theories with 
suitable compactifications and moduli stabilizations, one can in principle 
calculate the corresponding gauge couplings and Yukawa couplings at the string scale, 
which should be required to be consistent with the low-energy experimental values via RGE running. 
In particular, gauge and Yukawa couplings 
are completely determined by the string compactifications and moduli 
stabilizations above the SUSY-breaking scale. Thus, they are not 
related to the naturalness of the MSSM, 
and then we will not consider their fine-tuning measures, in particular $\Delta_{\rm BG}(\lambda_t)$, 
for our phenomenological considerations. 
Furthermore, we see that even considering $\Delta_{\rm BG}(\lambda_t)$, we can find solutions with a reduced FT price taking smaller values of $\lambda_t \times \Lambda$. 
Model  {\bf A3} is an example of reduced sensitivity to $\lambda_t$, featuring Higgsino NLSP.

Finally, models  {\bf B1}-{\bf B3} represent interesting corners of the parameter space.
 {\bf B1} is an example of a scenario with a stau NLSP, while a Bino-like neutralino is the NLSP for point  {\bf B2}.
In both cases, the NLSP is long lived, given the large mediation scale $M$, as we will discuss in the next section.  {\bf B3} instead is a model
with low $M$, hence a fast-decaying (Higgsino) NLSP.

\section{LHC phenomenology}
\label{sec:lhc}
As is any scenario with gauge-mediated SUSY breaking, the collider phenomenology of our models 
crucially depends on nature and properties  of the NLSP, its life-time in particular. In fact, independently of the production mechanism, 
any cascade decay will end with the NLSP decaying into a light gravitino LSP, 
whose mass is in terms of our parameters  $m_{\widetilde G}= {\Lambda \times M}/({\sqrt{3} M_{Pl}})$, 
$M_{Pl} = 2.4\times 10^{18}$ GeV.

As we have seen, the NLSP is the lightest neutralino in most of our parameter space. 
The decay modes crucially depend on the size of gaugino and Higgsino components of the lightest neutralino:
$\widetilde{\chi}_1^0 = N_{11} \widetilde{B} + N_{12} \widetilde{W}^0+ N_{13} \widetilde{H}_d^0+ N_{14} \widetilde{H}_u^0$.
In terms of the entries $ N_{1k}$, the widths of the possible decay modes of a neutralino NLSP to gravitino read
\cite{Ambrosanio:1996jn,Dimopoulos:1996yq}:
\begin{align}
\label{eq:gammachi}
\Gamma ( \widetilde{\chi}^0_1 \to \widetilde{G}\, Z)  & \simeq\frac{m_{\widetilde{\chi}^0_1}^5}{48 \pi\, m_{\widetilde G}^2 M_{Pl}^2} 
\left(\left|N_{12} c_\theta - N_{11}s_\theta \right|^2 + \frac{1}{2}\left|N_{13} c_\beta - N_{14}s_\beta \right|^2\right)
 \!\times\! \left(1-\frac{m_{Z}^2}{m_{\widetilde{\chi}^0_1}^2}  \right)^4, \\
\label{eq:gammachih}
\Gamma ( \widetilde{\chi}^0_1 \to \widetilde{G}\, h )  & \simeq\frac{m_{\widetilde{\chi}^0_1}^5}{96 \pi\, m_{\widetilde G}^2 M_{Pl}^2} 
\left|N_{13} c_\beta + N_{14}s_\beta \right|^2
\left(1-\frac{m_{h}^2}{m_{\widetilde{\chi}^0_1}^2}  \right)^4, \\
\Gamma ( \widetilde{\chi}^0_1 \to \widetilde{G}\, \gamma )  & \simeq\frac{m_{\widetilde{\chi}^0_1}^5}{48 \pi\, m_{\widetilde G}^2 M_{Pl}^2} 
\left|N_{11} c_\theta + N_{12} s_\theta \right|^2, 
\label{eq:gammachigamma}
\end{align}
where $s_\theta\equiv \sin\theta_W$, $c_\theta\equiv \cos\theta_W$, and  $s_\beta\equiv \sin\beta$, $c_\beta\equiv \cos\beta$,
and the dependence of the phase space on the gravitino mass has been neglected. 
As we can see from these equations, a mostly Higgsino NLSP will decay to $Z$ and $h$, while a sizable branching ratio to $\gamma$ is only possible if  $\widetilde{\chi}^0_1$ is mostly gaugino, e.g.~$\widetilde{B}$.

According to our scan, the only other possible NLSP is the lightest stau. In such a case, the decay width is:
\begin{equation}
\label{eq:gammatau}
\Gamma ( \widetilde{\tau}_1 \to \widetilde{G}\, \tau ) \simeq \frac{m_{\widetilde{\tau}_1}^5}{48 \pi\, m_{\widetilde G}^2 M_{Pl}^2},
\end{equation}
\begin{figure}[t]
\begin{center}
\includegraphics[width=0.6\textwidth]{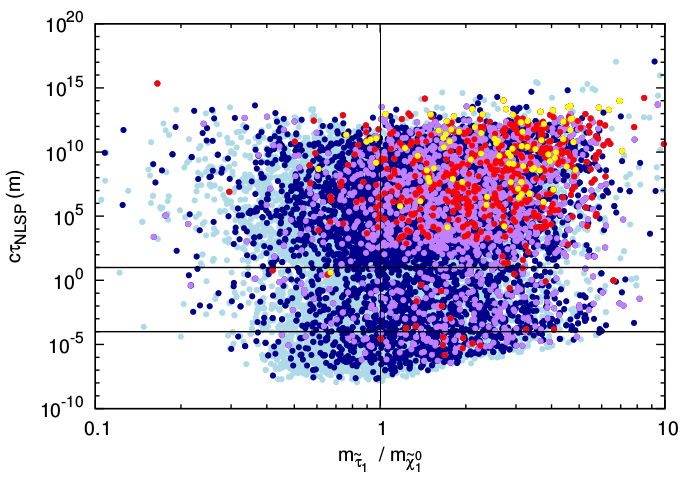}
\caption{The NLSP decay length vs.~the stau-neutralino mass ratio, which manifests the NLSP nature. Colors show
different ranges of $\Delta_{\rm EW}$ as in Fig.~\ref{fig:spectrum1}.\label{fig:ctau} }
\end{center}
\end{figure}

The expressions in Eqs.(\ref{eq:gammachi}-\ref{eq:gammatau}) show that the NLSP decay rate is 
always inversely proportional to the gravitino mass, hence on the gauge mediation scale, as $m_{\widetilde G}\propto M$. Therefore,
higher messenger scales will correspond to more long-lived NLSP.
We employ these formulae to compute the NLSP decay length $c\tau_{\rm NLSP} \equiv  c / \Gamma^{\rm tot}_{\rm NLSP}$ for the points of our scan. The result is shown in 
Fig.~\ref{fig:ctau}. The horizontal lines correspond to $c\tau_{\rm NLSP} = 0.1$ mm and 10 m. Points below  $c\tau_{\rm NLSP} = 0.1$ mm feature a NLSP decay that will mostly appear at the LHC as occurring promptly at the $pp$ collision point, while points above $c\tau_{\rm NLSP} = 10$ m likely give raise to NLSP decays outside the detector. As we can see from the plot, 
solution with low fine tuning tend to have a long-lived NLSP. This is because a large mediation scale, thus a long running, 
more easily achieve a partial cancellation among the terms in Eq.~(\ref{eq:rge}). Nevertheless, we see that some points with 
$\Delta_{\rm EW} < 100$ correspond to a promptly-decaying neutralino NLSP. 

\begin{figure}[t]
\begin{center}
\includegraphics[width=0.47\textwidth]{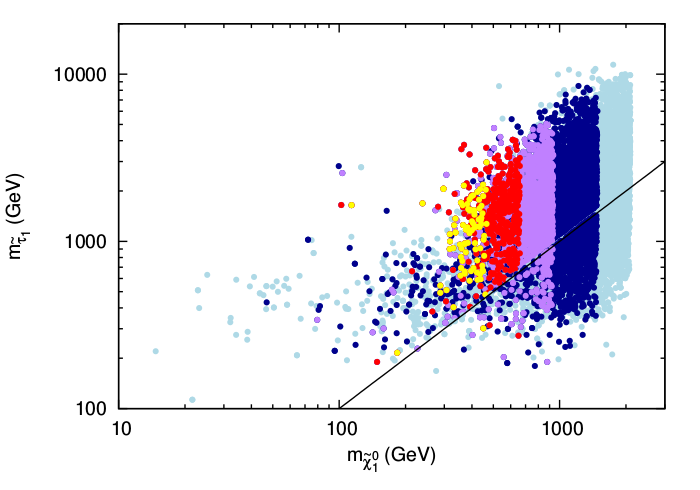}
\hfill
\includegraphics[width=0.47\textwidth]{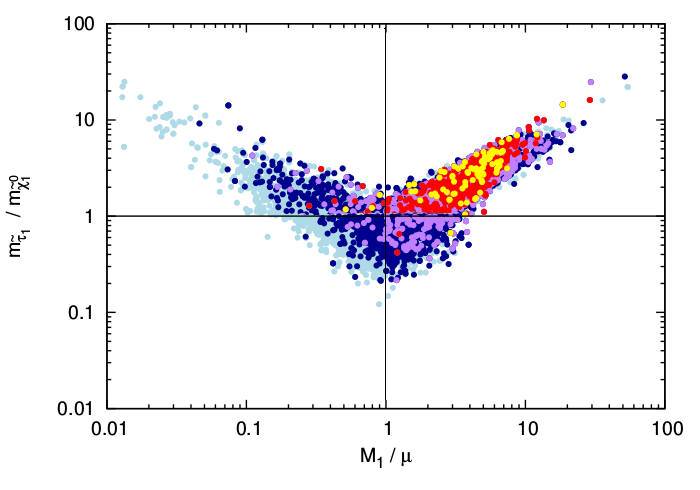}
\caption{The same as Fig.~\ref{fig:spectrum3} after imposing  the bounds in Eqs.~(\ref{eq:staubound1}, \ref{eq:staubound2}) for the points with a long-lived stau NLSP. \label{fig:spectrum4} }
\end{center}
\end{figure}

We are now ready to discuss phenomenology and constraints of our models at colliders, in particular at the LHC, according to the properties of the NLSP. 
\paragraph{Long-lived Higgsino NLSP.} 
As we have seen, our typical low $\Delta_{\rm EW}$ scenarios feature a small Higgsino mass and a large mediation scale, hence 
almost degenerate $\widetilde{\chi}^0_{1,2}$ and $\widetilde{\chi}^\pm_{1}$, with $\widetilde{\chi}^0_{1}$ 
as a long-lived NLSP. Models  {\bf A1}-{\bf A3} shown in Tab.~\ref{tab:benchmarks} belong to this class. 
If Higgsinos are the only light states, while the other SUSY particles such as stops and gluinos lie in the multi-TeV range, 
there is little room to test this case at the LHC. The cross section of the direct EW production
of $\widetilde{\chi}^0_{1,2}$ and $\widetilde{\chi}^\pm_{1}$ is rather large at the LHC with $\sqrt{s}=13$ TeV ($\approx34$~fb for $\mu=500$ GeV, summing together all modes \cite{resummino}) but $\widetilde{\chi}^0_1$ leaves the detector unseen, and the decay products of 
$\widetilde{\chi}^0_{2}$ and $\widetilde{\chi}^\pm_{1}$ are too soft to be observable at the LHC, due to the small mass gaps. 
Searches for events with large missing energy and a single jet (from initial state radiation) do not improve the situation
for Higgsinos heavier than 150$\div$200 GeV \cite{Han:2013usa,Gori:2013ala,Schwaller:2013baa,Baer:2014cua,Han:2014kaa,Baer:2014kya}.
However, this scenario is easily accessible at $e^+ e^-$ colliders with $\sqrt{s} \gtrsim 2\times m_{\widetilde{\chi}^\pm_{1}}$,
such as the International Linear Collider (ILC) \cite{Baer:2014yta}. From the scan, we found that $\widetilde{\chi}^0_{1} \lesssim 450$ (650) GeV for $\Delta_{\rm EW} < 50$ (100), cf.~Fig.~\ref{fig:spectrum3}, left. From this, we see that ILC operating 
at $\sqrt{s}$ up to about 1 TeV would test all our models with $\Delta_{\rm EW} < 50$, i.e.~tuning better than the 2\% level.
Another possibility relies on the production of heavier SUSY particles. As we have seen, the novel feature of models with matter-messenger couplings such as $\lambda_t$ (compared to our previous study \cite{Calibbi:2016qwt}) is the possibility of a light stop (and relatively lighter gluino), as illustrated by Fig.~\ref{fig:spectrum1} and the model  {\bf A2} of Tab.~\ref{tab:benchmarks}. 
This case is in principle testable at the LHC (relying on stop/gluino production) through searches for events with b-jets and missing energy aiming at standard `natural SUSY' scenarios. According to the recent study \cite{Buckley:2016kvr}, a spectrum like that of model  {\bf A2} already lies at the edge of the exclusion provided by the early 13 TeV LHC data, which approximately corresponds to $m_{\widetilde{t}_{1}}\gtrsim 1$ TeV for  $m_{\widetilde{g}}= 2$ TeV. Therefore, we expect that the present LHC run will start testing at least the bottom-left corner of the left plot in Fig.~\ref{fig:spectrum1}.
The possible presence of an intermediate Bino does not change the picture as the stop will always prefer to decay directly to Higgsinos, given the large $\sim y_t$ stop-Higgsino couplings.
\paragraph{Promptly-decaying Higgsino NLSP.} 
As we mentioned above, certain solutions, shown in the right-bottom corner of Fig.~\ref{fig:ctau}, have a (mostly Higgsino) neutralino,  with such a short lifetime that it would always appear to decay promptly at the collision point. An example is provided by 
the model  {\bf B3} in Tab.~\ref{tab:benchmarks}.
From Eqs.~(\ref{eq:gammachi}, \ref{eq:gammachih}), we see that an almost pure Higgsino mostly decays to $Z$ or $h$ and the gravitino, while the BR($\widetilde{\chi}^0_1\to \gamma \widetilde G$) is suppressed by the negligible gaugino component 
in $\widetilde{\chi}^0_1$. 
Furthermore, for moderate to large values of $\tan\beta$, the rates of the two modes $\widetilde{\chi}^0_1\to Z \widetilde G$
and $\widetilde{\chi}^0_1\to h \widetilde G$ only differ by the phase space, typically giving 
${\rm BR}(\widetilde{\chi}^0_1\to Z \widetilde G) \simeq 60\div 50\,\%$, ${\rm BR}(\widetilde{\chi}^0_1\to h \widetilde G) \simeq 40\div 50\,\%$,
for the range of $m_{\widetilde{\chi}^0_1}$ we are interested in.
A search for Higgsinos promptly decaying to a light gravitino employing the 8 TeV LHC data
has been published in \cite{Khachatryan:2014mma}. 
The limit is given as a function of ${\rm BR}(\widetilde{\chi}^0_1\to h \widetilde G)$ and reads $m_{\widetilde{\chi}^0_1} \gtrsim 325~(300)$ GeV for ${\rm BR}(\widetilde{\chi}^0_1\to h \widetilde G) = 40~(50)\,\%$. This constraint is too weak to exclude
our models with $\Delta_{\rm EW} < 100$, but we can expect a substantial improvement from the data collected at 13 TeV.
\paragraph{Long-lived stau NLSP.} 
Models with a long-lived stau (hence charged) NLSP can be much more easily tested at the LHC through searches for charged
tracks. The points in the top-left sector of Fig.~\ref{fig:ctau}, in particular the model  {\bf B1} of Tab.~\ref{tab:benchmarks}, belong  to this scenario. This model is already excluded by a CMS search with 8 TeV data for long-lived massive particles \cite{Chatrchyan:2013oca}, according to which the bound on the stau mass for staus directly produced through the Drell-Yan mechanism is:
\begin{equation}
\label{eq:staubound1}
m_{\widetilde{\tau}_1}>339~{\rm GeV}.
\end{equation}
This limit has not been improved yet by 13 TeV data \cite{Khachatryan:2016sfv}. 
This bound obviously becomes stronger if the production cross section of the heavier particles (e.g.~charged and neutral Higgsinos in our case) is larger than the stau production. In fact, any SUSY event will eventually feature two charged tracks at the end of the cascade decay. Assuming that the acceptance of events from such cascade decays is the same as in the case of direct stau production,\footnote{This assumption is corroborated by the fact that the limits reported in \cite{Chatrchyan:2013oca} on the production cross section for both direct and indirect (cascade decay) production almost overlap.} 
we can approximately recast the CMS bound by simply requiring that the production cross section of the heavier SUSY particles (that we computed by means of {\tt PROSPINO} \cite{Beenakker:1996ed}) does not exceed the Drell-Yann cross section of a mostly RH stau with $m_{\widetilde{\tau}_1}=339$ GeV: $\sigma_{\rm prod} \lesssim$ 0.32 fb. Such bound translates on the following limits on the relevant sparticle masses, valid for all models with a long-lived stau NLSP:
 \begin{align}
m_{\widetilde{\ell}_R} &\gtrsim 390~{\rm GeV},&
|\mu| &\gtrsim 840~{\rm GeV},  \nonumber \\
m_{\widetilde{t}_{1}} &\gtrsim 1~{\rm TeV},& 
m_{\widetilde{g}} &\gtrsim 1.5~{\rm TeV},
\label{eq:staubound2}
\end{align}
 where the bound on $m_{\widetilde{\ell}_R}$ refers to mass-degenerate RH selectron and smuon, and the bound on $|\mu|$ 
 comes from all possible combinations of production involving neutral and charged Higgsinos. This latter constraint is particularly stringent, and excludes all our points with $\Delta_{\rm EW}< 100$ featuring a long-lived stau NLSP.\footnote{If instead only the bound from direct stau production, Eq.~(\ref{eq:staubound1}), is applied, only about one fourth of the solutions with 
 $\Delta_{\rm EW}< 100$ are excluded.}
 This can be seen comparing Fig.~\ref{fig:spectrum4}, where the above bounds were applied, with the previous  Fig.~\ref{fig:spectrum3}.
Thus, searches for charged tracks already provide a stringent constraint on natural models belonging to this class, leaving
as viable solutions only models with a tuning more severe than 1\%.
\paragraph{Long-lived Bino NLSP.} 
This case is represented by the points in the top-left quadrant of Fig.~\ref{fig:spectrum3}.
In our case, models with a Bino-like neutralino NLSP resemble the ordinary searches for SUSY in gravity mediation, as the neutralino is long lived and decays outside the detector, appearing as missing energy. In our scenario several particles need to be light if Bino is light: those whose masses are controlled by $b^M_1$ in Eqs.~(\ref{eq:gauginos}, \ref{eq:scalars}), i.e.~the RH stau and sleptons, plus of course the Higgsinos in the case of low-tuned models. 
This is the case of our model  {\bf B2}. Therefore, there is room to test models of this kind by means of the EW production of these particles and the following cascade decays. The most promising channels
rely on production of Higgsinos, that will decay dominantly to staus (given the hierarchy of the lepton Yukawas) or directly to $ \widetilde{\chi}^0_{1}$ in the low $\tan\beta$ regime, as well as production of first and second generation RH sleptons:
\begin{align*}
& pp ~\to~ \widetilde{\chi}^+_{1} \widetilde{\chi}^-_{1} ~\to~ \widetilde{\tau}^+_{1} \widetilde{\tau}^-_{1} \nu_\tau \accentset{-}{\nu}_\tau ~\to~ \tau^+\tau^- + \nu_\tau \accentset{-}{\nu}_\tau +2\, \widetilde{\chi}^0_{1} \\
& pp ~\to~ \widetilde{\chi}^\pm_{1} \widetilde{\chi}^0_{2,3} ~\to~ \widetilde{\tau}^\pm_{1} \widetilde{\tau}^\pm_{1} \tau^\mp \accentset{(-)}{\nu}_\tau ~\to~\tau^\pm \tau^+\tau^-  +\accentset{(-)}{\nu}_\tau +2\, \widetilde{\chi}^0_{1} \\
& pp ~\to~  \widetilde{\chi}^0_{2} \widetilde{\chi}^0_{3} ~\to~ 
\widetilde{\tau}^\pm_{1} \tau^\mp \widetilde{\tau}^\pm_{1} \tau^\mp ~\to~ \tau^+\tau^-\tau^+\tau^-+2\, \widetilde{\chi}^0_{1} \\
& pp ~\to~ \widetilde{\chi}^\pm_{1} \widetilde{\chi}^0_{2,3} ~\to~ W^\pm Z + 2\, \widetilde{\chi}^0_{1}\\
& pp ~\to~ \widetilde{\ell}_R^+  \widetilde{\ell}_R^- \to \ell^+\ell^-+ 2\, \widetilde{\chi}^0_{1}
\end{align*}
All these modes have been intensively searched for by ATLAS and CMS with the 8 TeV data set. The limits on the Higgsino mass
reach up to 450 GeV, but quickly drop for $m_{\widetilde{\chi}^0_{1}}$ above 100$\div$150 GeV or small mass splittings \cite{Aad:2014nua,Aad:2014vma,Khachatryan:2014qwa,Aad:2014yka,Khachatryan:2016trj}, hence have no impact on the parameter space of our models. Nevertheless, scenarios with much heavier neutralinos will be accessible at the high-luminosity runs of the LHC, as demonstrated by e.g.~a prospect study for direct stau pair production \cite{ATL-PHYS-PUB-2016-021}.
\paragraph{NLSP decaying inside the detector.} 
As shown in Fig.~\ref{fig:ctau}, we found some low-tuned models ($\Delta_{\rm EW} < 100$) with
$0.1~{\rm mm}< c\tau_{\rm NLSP} < 10~{\rm m}$. In this regime the NLSP is likely to decay 
inside the detector after traveling a finite distance.
Most of the solutions of this kind feature a Higgsino NLSP that, as we have seen above, decays with comparable probabilities to $\widetilde G Z$ or $\widetilde G h$, cf.~Eqs.~(\ref{eq:gammachi}, \ref{eq:gammachih}). 
Possible strategies to test this kind of displaced neutralino decays at the LHC have been proposed in \cite{Meade:2010ji}.
A search for this kind of topology in association with jets -- thus sensitive to the production of colored superpartners -- has been published by ATLAS employing the 8 TeV run data \cite{Aad:2015rba}. Assuming BR($ \widetilde{\chi}^0_{1}\to \widetilde G h$)=100\%, they set a limit on the production cross section up to 1 fb for $10~{\rm mm} \lesssim c\tau_{\widetilde{\chi}^0_{1}} \lesssim 100~{\rm mm}$, corresponding to a gluino as heavy as $\sim$1.4 TeV. Our solutions with a Higgsino NLSP in this regime feature heavier gluinos $\gtrsim$ 2 TeV, but this shows that searches for displaced vertices may become sensitive to this special class of models in the future.

From Fig.~\ref{fig:ctau} we can also see that a few low-tuned models feature a stau NSLP with life-time in this intermediate regime.
Since the stau can only decay to $\widetilde G \tau$, the characteristic signature would be a charged track followed 
by a tau, i.e.~by a lighter lepton or a jet. Although triggering on these daughter particles at the LHC seems to be unfeasible, 
searches for disappearing tracks \cite{Aad:2013yna,CMS:2014gxa} are sensitive to such a scenario, as shown in \cite{Evans:2016zau}, maximally for $c\tau_{\widetilde{\tau}_1} \approx 50$ cm. Furthermore, for $c\tau_{\widetilde{\tau}_1} \gtrsim 2$ m, a substantial fraction of the staus decay outside the detector such that the searches for stable charged particles discussed above 
become increasingly sensitive \cite{Evans:2016zau}. This seems to be the best way to test the few models of this kind we found, 
that have $m_{\widetilde{\tau}_1}\approx 300$ GeV  and $c\tau_{\widetilde{\tau}_1} \gtrsim 3$ m.

\section{Summary and discussion}
\label{sec:conclusions}
Within the MSSM with gauge-mediated SUSY breaking, 
we have discussed a class of models characterized by two deviations from minimal setups: 
(i) a non-unified messenger sector providing large freedom in the gaugino mass ratios, and (ii) a matter-messenger
coupling $\lambda_t$ inducing 1-loop A-terms and additional contributions to the scalar masses. 
The first ingredient allows for a compensation in the running of the Higgs soft mass that reduces its sensitivity on the stop and
gluino masses, thus reducing the fine tuning. The best solutions were found for $M_2/M_3 \approx 3$ at the mediation 
scale. The second ingredient generalizes our previous work \cite{Calibbi:2016qwt} and gives the possibility of building models with 
a lighter stop (due to some negative contributions to its mass, as well as the impact on the Higgs mass of the $\lambda_t$-induced $A_t$) and FT as low as $\Delta_{\rm EW}\approx 20$, i.e.~around 5\%, while in setups with $\lambda_t = 0$ no solutions with a tuning better than about 2\% were found.

The typical spectra of the models with best FT that we found feature light Higgsinos and possibly light Bino and RH sleptons (in particular the lightest stau), since the masses of these particles receive no radiative contribution at one loop from the Wino mass that instead is required to be very heavy to trigger the compensation in the running that we discussed above. Other particles are typically heavy, with the possible exception of the lightest stop that can be as light as 1 TeV. Gluinos and RH squarks lie above 2 TeV, while LH squarks and sleptons are multi-TeV again because of the large Wino mass.
Therefore, this class of models is a remarkable example of SUSY scenarios with low fine tuning but spectra that can be easily evade
current LHC searches,\footnote{A notable exception is provided by models featuring a long-lived stau NLSP that, as we have seen in the previous section, are already excluded down to FT$\approx$1\%.} 
in contrast to the classical `natural SUSY' framework whose status has been recently studied in \cite{Buckley:2016kvr}. 
This feature is shared by other SUSY setups, for instance in the context of gravity mediation by `radiatively natural' models (for
recent discussions see \cite{Baer:2016usl,Baer:2016wkz}). 
Our framework can be in principle distinguished by this latter one, because it features lighter gluinos, some lighter scalars, and have instead heavy Winos.
Besides these differences, we also find that light higgsinos is a minimal condition for low tuning, while other SUSY particles are not necessarily light. Approximately we have 5\% (2\%) FT for 300 (450) GeV Higgsinos. This makes a strong case
for the ILC, that could in fact test several classes of models with tuning better than 2\%, if operating at the center of mass
up to 1 TeV. This is an important conclusion that has been reached elsewhere in the literature, see e.g.~\cite{Baer:2016new}, and we want to remark it here.
Possible exceptions to the above conclusion 
could be given by models with a high degree of correlations among the high energy parameters,
such as in `supernatural' SUSY scenarios in the context of no-scale supergravity \cite{Leggett:2014mza,Leggett:2014hha,Du:2015una,Li:2015dil}, or by models with non-minimal Higgs sectors, e.g. resembling N=2 SUSY \cite{Ding:2015epa}.
In these cases low tuning can be achieved even for heavy Higgsinos.

\section*{Acknowledgments}
This research was supported in part by the 
Natural Science Foundation of China under 
grant numbers 11135003, 11275246, and 11475238 (TL),
and the CAS-TWAS President's Fellowship Programme (WA).

\end{document}